\documentclass[12pt]{article}
\usepackage{graphicx}
\usepackage{cite}
\usepackage{color}
\newcommand{\bm}[1]{\mbox{\boldmath $#1$}}
\newcommand{\fnd}[2]{\frac{\textstyle #1}{\textstyle #2}}
\newcommand{\x}[1]{{\textstyle #1}}
\newcommand{\abs}[1]{\left| #1\right|}

\newcommand{\xrm}[1]{{\textstyle \mbox{\rm #1}}}

\newcommand{\Imag}[1]{\Im {\it m}\left(#1 \right)}

\begin{document}
\title{
Deducing the string-breaking distance\\ in strong production processes}
\author{
Eef van Beveren$^{1}$ and George Rupp$^{2}$\\ [10pt]
{\small\it $^{1}$Centro de F\'{\i}sica Te\'{o}rica,
Departamento de F\'{\i}sica,}\\ {\small\it Universidade de Coimbra,
P-3004-516 Coimbra, Portugal}\\ {\small\it http://cft.fis.uc.pt/eef}\\ [10pt]
{\small\it $^{2}$Centro de F\'{\i}sica das Interac\c{c}\~{o}es Fundamentais,
Instituto Superior T\'{e}cnico,}\\
{\small\it Universidade T\'{e}cnica de Lisboa, Edif\'{\i}cio Ci\^{e}ncia,
P-1049-001 Lisboa, Portugal}\\ {\small\it george@ist.utl.pt}\\ [10pt]
{\small PACS number(s): 11.80.Gw, 11.55.Ds, 13.75.Lb, 12.39.Pn}
}


\maketitle

\begin{abstract}
We show that
the string-breaking distance can be read off from meson-production data,
by employing a previously derived expression for the production amplitude.
Accordingly, we find that the radii of 0.67, 0.34 and 0.20 fm
for the creation of non-strange $q\bar{q}$ pairs
obtained in the Resonance-Spectrum-Expansion model,
for light-quark, $c\bar{c}$, and $b\bar{b}$ environments, respectively,
are in perfect agreement with $S$-wave di-pion production data,
upon employing an ansatz with no additional free parameters.
\end{abstract}

Ever since the reconciliation of quark confinement and hadronic decay
by G.~Zweig's proposal \cite{CERNREPTH401/412}, quark-pair ($q\bar{q}$)
creation and annihilation constitute an important piece of research in
particle physics.
Later, one realised, as applied in hadron models \cite{ZPC21p291},
that the created $q\bar{q}$ pairs are favored to carry
the quantum numbers of the vacuum \cite{PRD7p2136,PRD2p336},
which is confirmed by the experimentally observed
dominant hadronic decay modes \cite{JPG33p1}.
Inspiration from the theory of strong interactions
has given rise to the flux-tube picture \cite{PLB124p247}
and the gluonic string model \cite{NPA518p358,PLB228p167}.

In the string picture, the gluonic field energy increases
with larger interquark distances.
Hence, the energy needed for the creation of a $q\bar{q}$ pair
depends on the distance between the hadron constituents.
Estimates for the average distance at which $q\bar{q}$ pair creation takes
place yield, for mesons, values ranging from a few tenths of
\cite{PRD71p114513} to several \cite{NPB414p815} fermis.
Moreover, recently the shape of the transition potentials
derived in Ref.~\cite{ZPC21p291}
has been confirmed by lattice calculations  \cite{PRD71p114513}.

In the Resonance-Spectrum-Expansion meson model \cite{PRD27p1527},
the field energy $\varepsilon$ stems from the confinement potential,
which is quadratic in distance
and has a flavor-independent level spacing $\omega$.
As a consequence, considering a meson with constituents of flavor f,
the average string-breaking distance $r_{0,\,\xrm{\footnotesize f}}$
for the creation of a light $q\bar{q}$ pair is inversely proportional
to the square-root \cite{PRD21p772}
of the reduced mass $\mu_\xrm{\footnotesize f}$
of the meson constituents, i.e.,
\begin{equation}
\mu_\xrm{\footnotesize f}\,
r_{0,\,\xrm{\footnotesize f}}^{2}\; =\;
\fnd{2\varepsilon_{q\bar{q}}}{\omega^2} \; ,
\;\;\;\xrm{where}\;\;\;
q=u,d,\,\xrm{or}\; s \;.
\label{r0}
\end{equation}
Here, we employ the flavor masses of Ref.~\cite{PRD27p1527}
($m_{n}\equiv m_{u/d}=406$, $m_{s}=508$, $m_{c}=1562$, and $m_{b}=4724$ MeV)
and $r_{0,n\bar{s}}=3.24$ GeV$^{-1}$ \cite{PLB641p265}.
Hence, with $n\bar{n}$, $c\bar{c}$, and $b\bar{b}$
for the constituents of a decaying meson,
we obtain $r_{0,n\bar{n}}=0.67$ fm, $r_{0,c\bar{c}}=0.34$ fm,
and $r_{0,b\bar{b}}=0.20$ fm, respectively,
for the creation of a light quark pair.
Below, we shall confront these values with the data.

In Refs.~\cite{ARXIV07064119,ARXIV07105823}
we have discussed the expression \cite{footnote}
\begin{equation}
\bm{P}\; =\;\Imag{\bm{Z}}\; +\; T\,\bm{Z}
\; ,
\label{Production}
\end{equation}
relating the two-body subamplitude \bm{P},
showing up in processes of strong decay under the spectator assumption,
with elements of the two-body scattering amplitude $T$.
The vector \bm{Z} represents the quark-pair creation vertex,
and does not carry information on the two-body
final-state interactions, which are contained in $T$.

For complex coefficients $Z=\abs{Z}\exp (i\phi)$, we obtain
from Eq.~(\ref{Production}), in the one-channel case
(i.e., below the first inelastic threshold) and writing
$T=\left\{\exp (2i\delta )-1\right\} /2i$,
\begin{eqnarray}
P & = & \abs{Z}\left[
\frac{1}{2i}\left\{ e^\x{i\phi}-e^\x{-i\phi}\right\} +
\frac{1}{2i}\left\{ e^\x{2i\delta}-1\right\} e^\x{i\phi}
\right]
\nonumber\\ [3pt] & = & \abs{Z}\,e^\x{-i\phi}\,\frac{1}{2i}
\left\{ e^\x{2i(\delta +\phi )}-1\right\}
\; .
\label{proWatson}
\end{eqnarray}
A comparable result has recently also been advertised by D.V.~Bugg
\cite{EPJC54p73}, thereby distinguishing between $T(\xrm{elastic})$
and  $T(\xrm{production})$,
which here read $\left\{\exp (2i\delta )-1\right\} /2i$
and $\left\{\exp (2i(\delta +\phi ))-1\right\} /2i$, respectively.

Equation~(\ref{proWatson}) does seem not to
satisfy Watson's theorem \cite{PR88p1163}, which
states that the phase motion of elastic scattering $T$
and production $P$ should be the same \cite{PRD35p1633}.
However, in the following we shall find that the complex phase
of two-body subamplitude (\ref{proWatson})
is identical to the complex phase of $T$.

Assuming OZI-allowed strong processes,
the nonstrange ($n\bar{n}$) and strange ($s\bar{s}$)
scalar ($J^{PC}\!=\!0^{++}$) isoscalar ($I=0$) $q\bar{q}$ systems
couple dominantly to $\pi\pi$ below the $K\bar{K}$ threshold.
Hence, we may, to a good approximation, consider uncoupled
$\pi\pi$ elastic scattering and production,
for total invariant two-pion masses up to about 1 GeV.
In the one-channel case,
formula~(\ref{proWatson}) becomes a rather simple expression.
We find for the $S$-wave
two-pion production and elastic scattering amplitudes
the relation \cite{ARXIV07064119}
\begin{equation}
P_{0}\;\propto\;
j_{0}\left( pr_{0}\right)\, +\,
i\, T_{0}\, h^{(1)}_{0}\left( pr_{0}\right)
\; ,
\label{P2pidef}
\end{equation}
which, by the definition of Bessel and Hankel functions,
and the introduction of the elastic scattering phase shift
$\delta_{0}=\log\left\{\left( 1+2iT_{0}\right) /2i\right\}$,
reduces to
\begin{equation}
P_{0}\;\propto\;
\fnd{1}{pr_{0}}\,\sin\left[\delta_{0} +pr_{0}\right]\,
e^\x{i\delta_{0}}
\; .
\label{P2pi}
\end{equation}
Consequently, we obtain for production a phase which equals the sum of
the elastic scattering phase shift and a term
depending on the linear momentum $p$ of the two-pion system.
Next, we shall apply relation~(\ref{P2pi}) to the central production
of pion pairs in $pp\to pp\pi\pi$.

\begin{figure}[htbp]
\begin{center}
\begin{tabular}{c}
\includegraphics[height=160pt, angle=0]{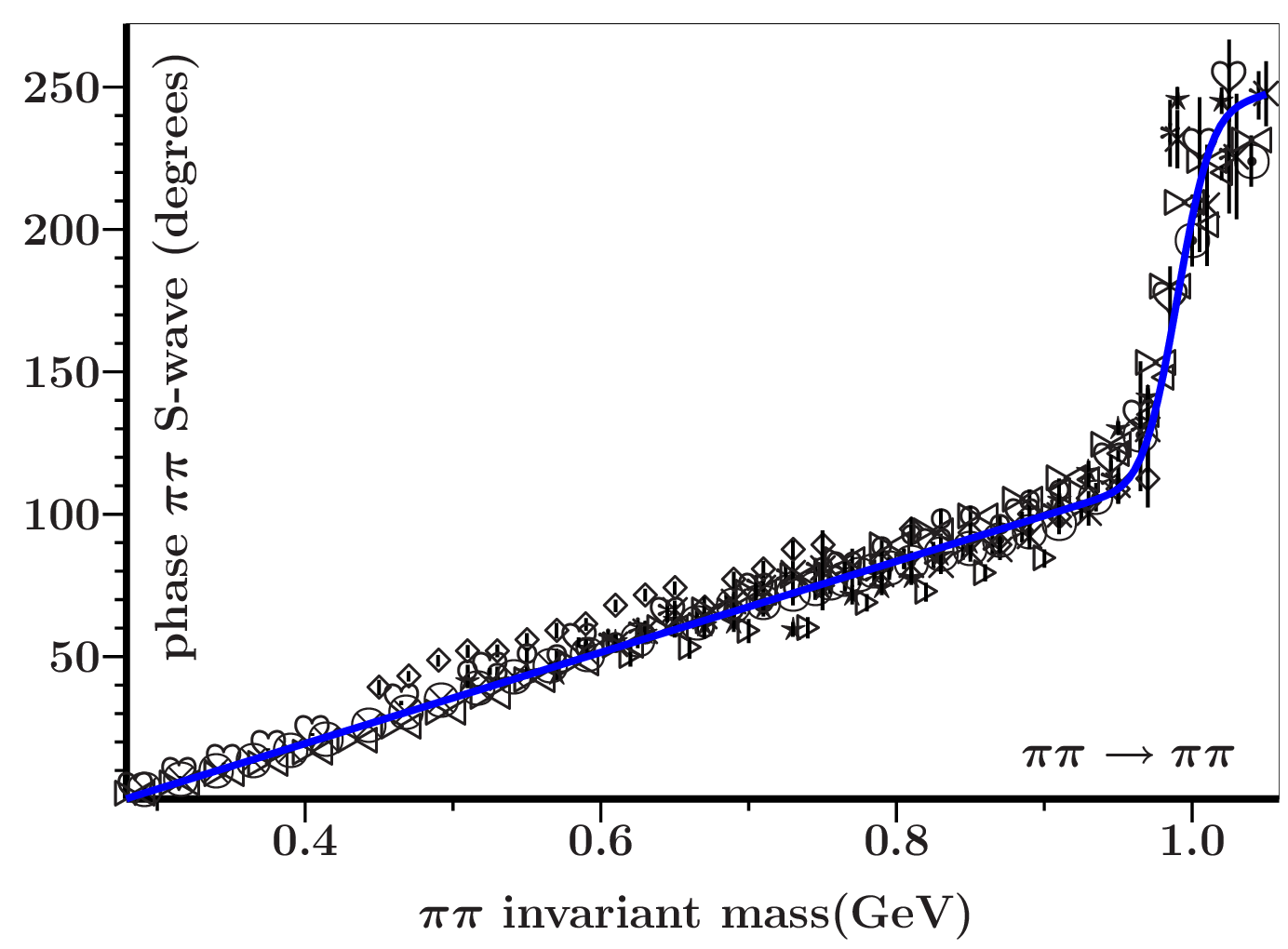}
\mbox{} \\ [-15pt]
\end{tabular}
\end{center}
\caption[]{
The various sets of elastic scattering data
used in this paper are taken from
(\bm{\odot}, Ref.~\cite{PRD7p1279});
(\bm{\ast}, Ref.~\cite{NPB64p134});
(\bm{\star},
\bm{\times},
\bm{\diamond},
\bm{\triangleleft},
\bm{\triangleright}
for analyses A, B, C, D, E of Ref.~\cite{NPB75p189}, respectively);
(\bm{\circ}, Ref.~\cite{NPB79p301});
(\bm{\otimes}, Ref.~\cite{PLB641p265});
(\bm{\bowtie}, Ref.~\cite{EPJC52p55});
(\bm{\heartsuit}, Ref.~\cite{PRL96p132001}).
The solid line indicates the average behavior of the data
as a function of the total invariant two-pion mass.
}
\label{elphases}
\end{figure}
For elastic $\pi\pi$ scattering, we use the analyses of
Refs.~\cite{PRD7p1279,NPB64p134,NPB75p189,
NPB79p301,EPJC52p55,PRL96p132001}.
Furthermore,
employing conservative estimates for the theoretical uncertainties,
we also use the theoretical phases of Ref.~\cite{PLB641p265},
which agree well with the data
presented by the BES collaboration in Ref.~\cite{PLB598p149}.
In Fig.~\ref{elphases} we depict the available experimental phase shifts.
The corresponding cross sections, being proportional to 
$\abs{\exp\left( 2i\delta_{0}\right)-1}^{2}/p^2$
below the $K\bar{K}$ threshold, peak around 600 MeV.

Using the recipe~(\ref{P2pi}), it is extremely easy to determine
the production amplitude that follows for the various
experimental phases $\delta_{0}$, once we have at our disposal a reasonable
value for the average distance $r_{0}$ at which quark-pair creation and
annihilation is supposed to occur.
Such an estimate was given long ago in Ref.~\cite{ZPC30p615},
and recently in Ref.~\cite{PLB641p265}.
The estimates in Refs~\cite{ZPC30p615,PLB641p265}
have been confirmed by recent lattice calculations \cite{PRD71p114513}, i.e.,
for light-quark-pair creation in heavy-quark systems.
For light consituents, we shall here use the most recent value of
$r_{0}=0.67$~fm used in Ref.~\cite{PLB641p265}.
\begin{figure}[htbp]
\begin{center}
\begin{tabular}{ccc}
\includegraphics[height=180pt, angle=0]{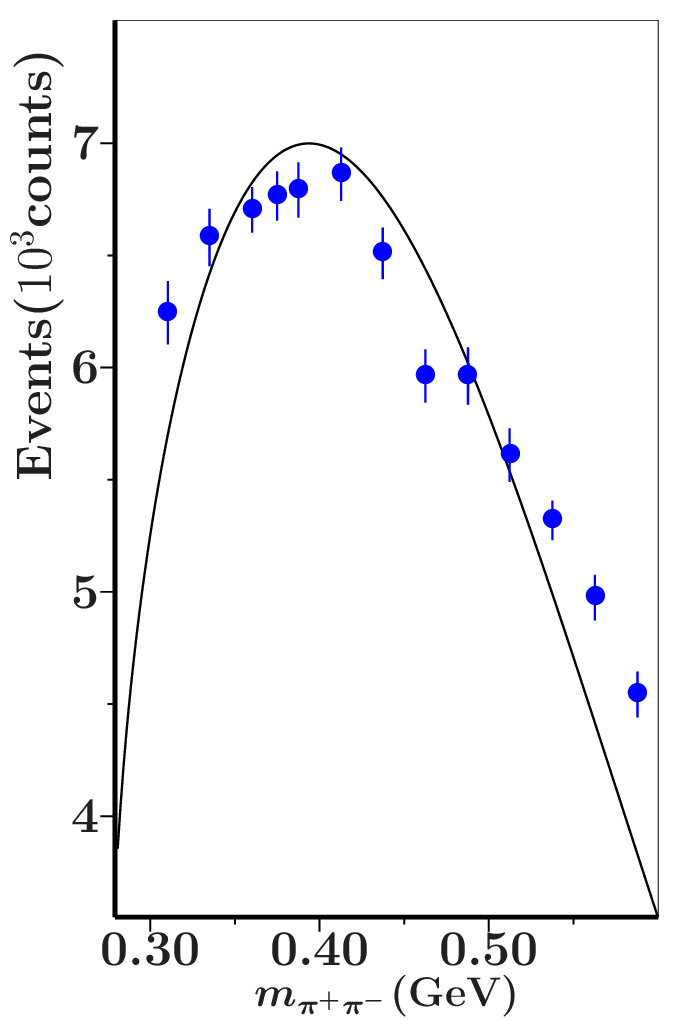} &
\resizebox{!}{180pt}{\includegraphics{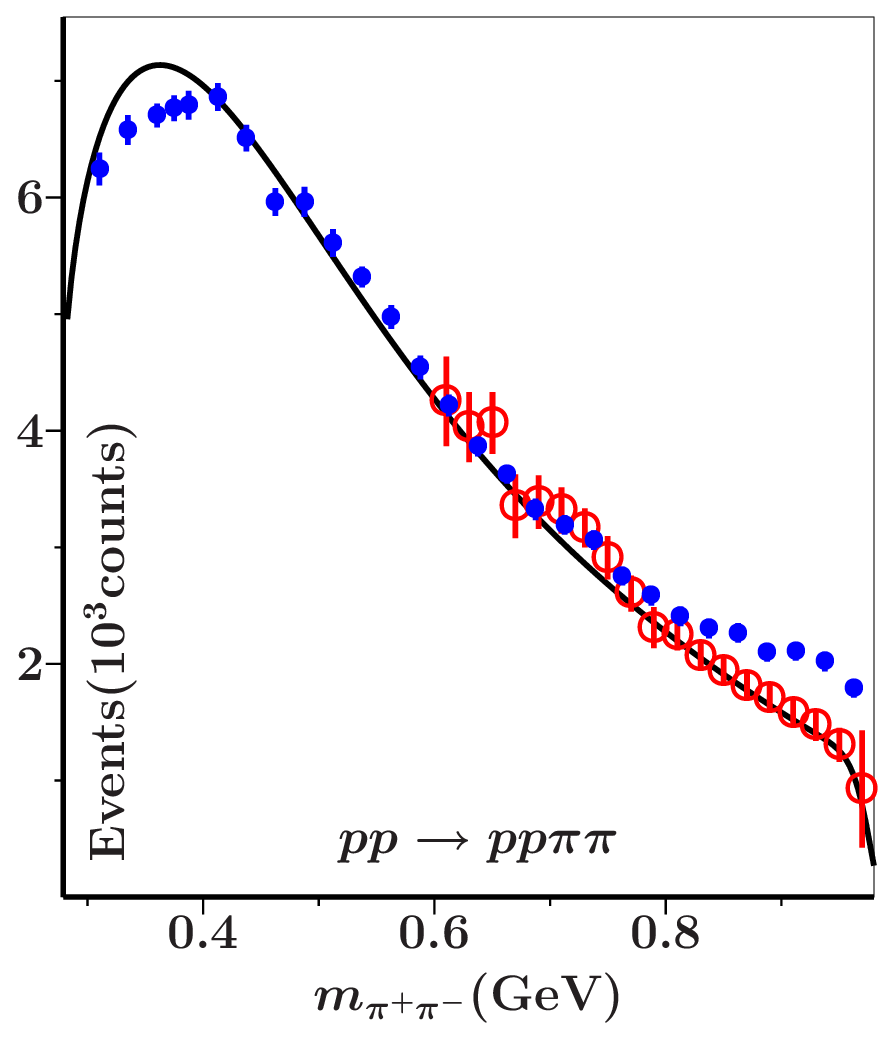}} &
\resizebox{!}{180pt}{\includegraphics{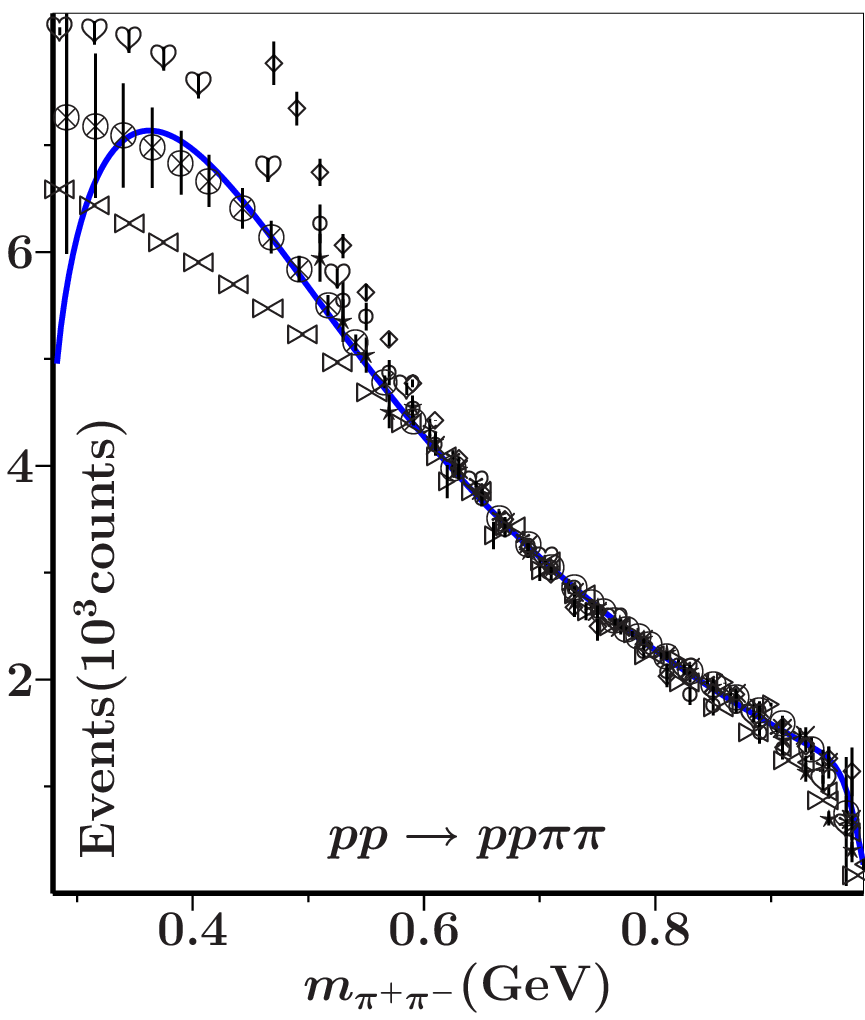}}\\ [-70pt]
\hspace{8pt}{\bf (a)} & \hspace{8pt}{\bf (b)} &
\hspace{8pt}{\bf (c)} \\ [45pt]
\end{tabular}
\end{center}
\caption[]{
\small (a) Expression (\ref{P2pi}) with $r_{0}=0.67$ fm
\cite{PLB641p265}, for the eye-guiding curve of Fig.~\ref{elphases} (black),
compared to the AFS data \cite{NPB264p154} (blue).
The normalization of the curve has been adjusted to the data. 
(b) Expression (\ref{Pexp2pi}) for the eye-guiding curve,
compared to the AFS data \cite{NPB264p154} ({\color{blue}$\bullet$}), and
to a production amplitude derived from a subset
of the elastic scattering data ({\color{red}\large $\circ$})
by D.~V.~Bugg \cite{EPJC54p73}, using a different method.
(c) Expression (\ref{Pexp2pi}) applied directly to the data
exhibited in Fig.~\ref{elphases} (black), to be compared with the
same production data as shown in (b).
}
\label{prodphases}
\end{figure}

Figure~\ref{prodphases}a shows how $\abs{P_{0}}^{2}$
varies with the total invariant two-pion mass
when expression (\ref{P2pi}) is applied to
the eye-guiding curve of Fig.~\ref{elphases}.
As a first result, we observe that
for the production amplitude (\ref{P2pi}),
we obtain a maximum at about 400 MeV.
This agrees with the central $\pi\pi$ production data
of the AFS collaboration \cite{NPB264p154} taken at ISR (CERN),
and also with the data of the BES collaboration \cite{PLB598p149}.
Similar conclusions can be found in the work of Au, Morgan, and Pennington
\cite{PRD35p1633}.

Summarizing so far, with one parameter, $r_{0}$,
which had been determined two decades ago for the modeling of
elastic meson-meson scattering \cite{ZPC30p615,PRD27p1527},
we obtain a good result for two-particle production at low energies,
by using a formalism predicting for production
a phase motion that differs from the elastic phase
by a momentum-dependent function.

However, production amplitude~(\ref{P2pi})
deviates from the data for larger values of the invariant two-pion mass.
This result could easily have been anticipated,
since from Eq.~(\ref{P2pi}) one observes that at higher energies
the discrepancy with elastic scattering increases.
Here, we assume that the complex phase of the vertex,
represented by the vector \bm{Z} in Eq.~(\ref{Production}),
stems from the time delay necessary for the initially created quark pair
to escape from the interaction region.
Naively, we expect that at higher energies this supposed constant time delay 
will have a relatively larger effect.
At this stage, it is worthwhile
to mention that expression~(\ref{P2pi}) stems from the one-delta-shell
approximation to the peaked potential that describes the transitions
between $q\bar{q}$ and meson-meson configurations.
Now, the $\delta$-shell approximation is good enough for
the demonstration of certain aspects of the meson spectrum
\cite{PRD21p772,PRL91p012003,PRL97p202001},
but not for a detailed description of meson-meson scattering at
high energies.

More accurate transition potentials have been determined in
Ref.~\cite{ZPC21p291}. Their form, as a function of the interquark distance
$r$, is reasonably well described by \cite{CPC27p377}
\begin{equation}
V_\xrm{t}(r)\;\propto\;
\fnd{r}{r_{0}}\, e^\x{-\frac{1}{2}\left( r/r_{0}\right)^{2}}
\; .
\label{Vtrans}
\end{equation}
>From this shape, which strongly resembles the form of the
$q\bar{q}\leftrightarrow$ meson-meson transition rate determined on the
lattice (see Fig.~18 of Ref.~\cite{PRD71p114513}), we infer
\cite{Calogero}
that for the phases in production processes one has
\begin{equation}
P_{0}\;\propto\;
\fnd{1}{pr_{0}}\,
\sin\left[\delta_{0} +pr_{0}\, e^\x{-\left( pr_{0}\right)^{2}}\right]\;
e^\x{i\delta_{0}}
\; .
\label{Pexp2pi}
\end{equation}
Note that, with the latter expression, also the real production phase tends
towards the elastic-scattering phase as $p$ increases.
Figure~\ref{prodphases}b shows the behavior of expression~(\ref{Pexp2pi})
for $\pi\pi$ production, when applied to the eye-guiding line of
Fig.~\ref{elphases}, for invariant masses below the $K\bar{K}$ threshold.
Above about 600 MeV, we compare our semi-theoretical production amplitude,
besides the AFS data, also to an analysis based on standard methods,
though restricted to $I=0$ and $S$-waves only \cite{EPJC54p73}, just as in the
present paper. The slight discrepancy between the AFS data and our predictios
above 800 MeV may be explained by $I=2$ and $D$-wave contributions
\cite{AIPCP814p706}.

In Fig.~\ref{prodphases}c we show the result
for the procedure defined in Eq.~(\ref{Pexp2pi})
applied separately to each of the data points exhibited in
Fig.~\ref{elphases}.
The general trend is similar to that of the curve
shown in Fig.~\ref{prodphases}b.
However, at low energies we note that the procedure (\ref{Pexp2pi})
is very sensitive to the precise values of the elastic phases.

The theoretical maximum in the curve of Fig.~\ref{prodphases}b
at an invariant two-pion mass of 363 MeV is a consequence of
the choice $r_{0}=0.67$ fm. The latter value was determined in our
unitarized meson model \cite{ZPC30p615,PLB641p265}
for light-quark-pair creation in the presence of light quarks.
Through flavor invariance of the strong interactions according to
Eq.~(\ref{r0}), it is related to the average radius
of light-quark-pair creation in the presence of heavy quarks
\cite{PRD21p772}.

In the process $J/\psi\to\omega\pi\pi$, we must consider the presence
of a $c\bar{c}$ system.
This implies $r_{0}=0.34$ fm, which causes
the maximum of the corresponding amplitude to come out at 467 MeV.
This appears to be the reason why in the experiment of the BES
collaboration \cite{PLB598p149} a $\sigma$ signal at such a higher energy
was observed. Nevertheless, the $\sigma$ pole position, contained in the
$T$ matrix, which is not altered by Eq.~(\ref{Production}),
is the same as for the results of the AFS collaboration
and also for the elastic $\pi\pi$ amplitudes.
\begin{figure}[htbp]
\begin{center}
\begin{tabular}{cc}
\includegraphics[height=220pt, angle=0]{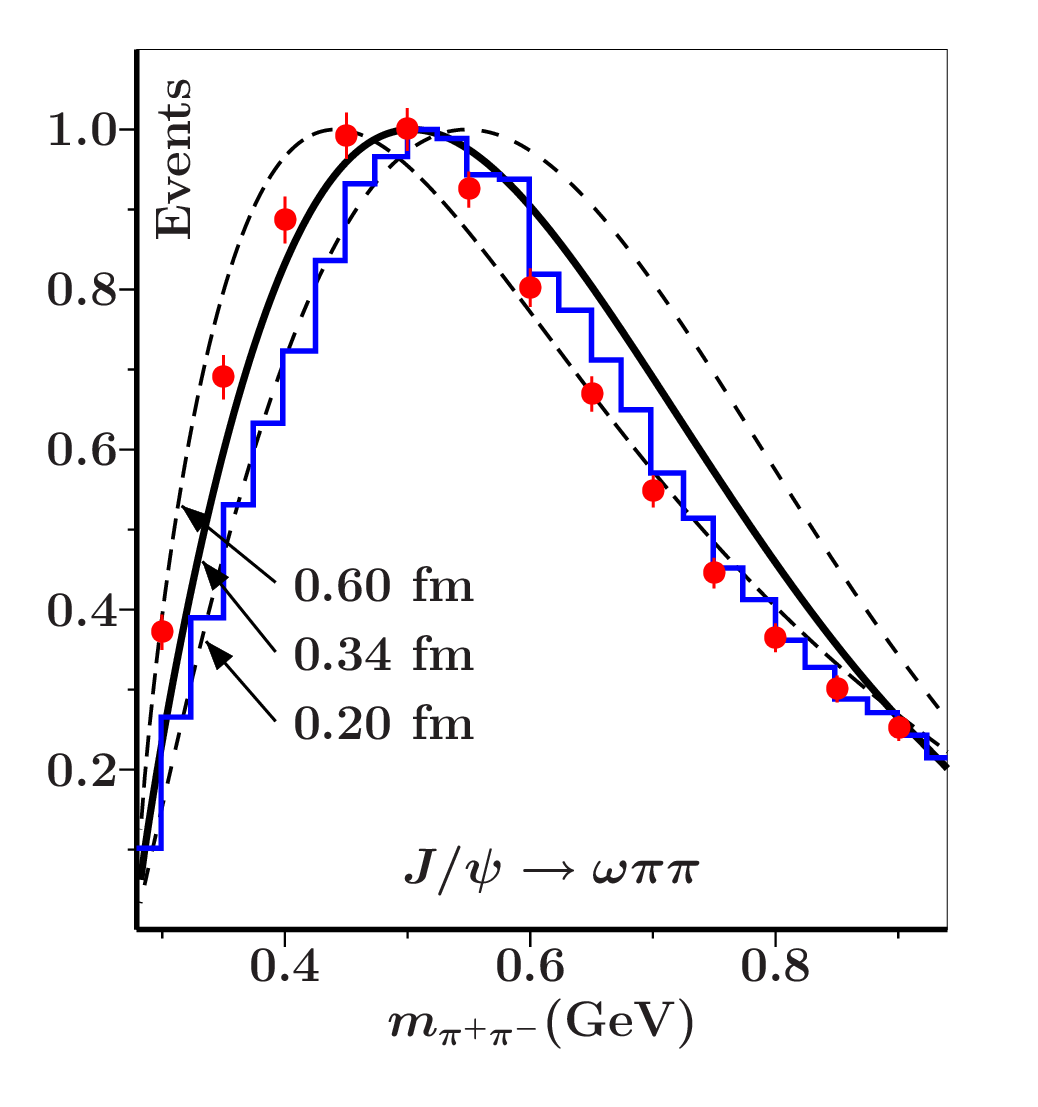} &
\resizebox{!}{220pt}{\includegraphics{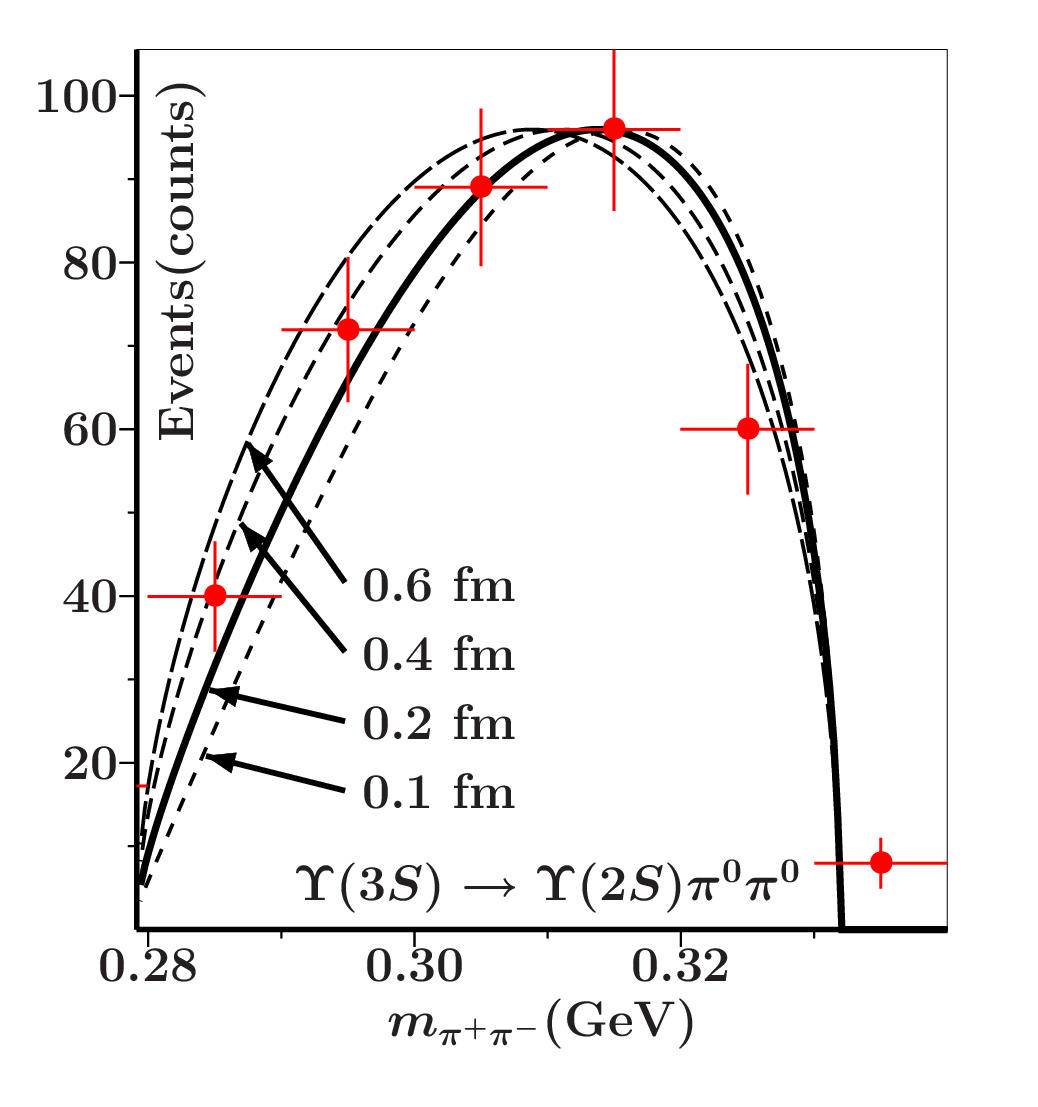}}\\ [-210pt]
\hspace{148pt}{\bf (a)} & \hspace{148pt}{\bf (b)} \\ [170pt]
\end{tabular}
\end{center}
\caption[]{\small
(a) Expression (\ref{Pexp2pi}) applied to the
eye-guiding curve, with $r_{0}=0.34$ fm \cite{PRD21p772}
(solid) and two different values of $r_{0}$ (dashed),
compared to
data of the BES collaboration \cite{PLB598p149} (blue histogram),
and to an analysis of the same data done by D.~V.~Bugg
in Ref.~\cite{EPJC52p55}, here multiplied with phase space
({\color{red}$\bullet$}).
Data and curves have been normalized to unity at the peak.
(b) As in (a), but now for $r_{0}=0.20$ fm
(solid) and also for different values of $r_{0}$ (dashed),
and compared to di-pion data ({\color{red}$\bullet$})
of the CLEO collaboration \cite{PRD76p072001}. 
}
\label{bescleo}
\end{figure}

In Fig.~\ref{bescleo}a we compare
the data of the BES collaboration \cite{PLB598p149},
and an analysis of the same data carried out by D.~V.~Bugg
in Ref.~\cite{EPJC52p55},
with the prediction of the procedure defined in Eq.~(\ref{Pexp2pi}),
for $r_{0}=0.34$ fm,
applied to the eye-guiding curve of Fig.~\ref{elphases}.
A similar comparison, but now for $r_{0}=0.20$ fm,
is given in Fig.~\ref{bescleo}b with the apparently clean $S$-wave di-pion
signal in $\Upsilon (3S)\to\Upsilon (2S)\pi^{0}\pi^{0}$
of the CLEO collaboration \cite{PRD76p072001}.
We again notice a nice agreement of our {\it parameter-free} \/procedure
with the data.  The production amplitudes, obtained from
Figs.~(\ref{bescleo}a,\ref{bescleo}b) by correcting for phase space, show
maxima at 467 MeV for the BES data and 521 MeV for the CLEO data.

In conclusion, our modification (\ref{Pexp2pi}) of
the elastic phases appears to explain well the differences
between $S$-wave production and elastic scattering data.
Furthermore, just above threshold, from the slope of the production curve
as a function of linear momentum, one can read from the data the interquark
distance at which pair creation is favored.
For light-quark-pair creation in the presence of light quarks,
we find $r_{0}=0.67$ fm, whereas in the presence of heavy quarks the
radii are $r_{0}=0.34$ fm for $c\bar{c}$ and
$r_{0}=0.20$ fm for $b\bar{b}$, the latter value being in
agreement with recent lattice calculations \cite{PRD71p114513}, too.

\section*{Acknowledgments}

We are indebted to D.~V.~Bugg for enlightening discussions,
to I.~J.~R.~Aitchison and I.~Caprini
for pertinent clarifications on Watson's theorem,
and also thank P.~Bicudo for drawing our attention to the lattice results in
Ref.~\cite{PRD71p114513}.
This work was supported in part by the {\it Funda\c{c}\~{a}o para a
Ci\^{e}ncia e a Tecnologia} \/of the {\it Minist\'{e}rio da Ci\^{e}ncia,
Tecnologia e Ensino Superior} \/of Portugal, under contract
PDCT/ FP/\-63907/\-2005.

\newcommand{\pubprt}[4]{{#1 {\bf #2}, #3 (#4)}}
\newcommand{\ertbid}[4]{[Erratum-ibid.~{#1 {\bf #2}, #3 (#4)}]}
\def\AIPCP{AIP Conf.\ Proc.}
\def\AP{Ann.\ Phys.}
\def\CPC{Comput.\ Phys.\ Commun.}
\def\EPJC{Eur.\ Phys.\ J.\ C}
\def\JPG{J.\ Phys.\ G}
\def\NPA{Nucl.\ Phys.\ A}
\def\NPB{Nucl.\ Phys.\ B}
\def\PLB{Phys.\ Lett.\ B}
\def\PR{Phys.\ Rev.}
\def\PRD{Phys.\ Rev.\ D}
\def\PRL{Phys.\ Rev.\ Lett.}
\def\ZPC{Z.\ Phys.\ C}

\end{document}